\documentclass[aps,prc,groupedaddress,showpacs,floatfix,11pt]{revtex4}

\usepackage{graphicx}
\usepackage[dvips]{color}
\usepackage{colordvi}

\def\gtorder{\mathrel{\raise.3ex\hbox{$>$}\mkern-14mu
 \lower0.6ex\hbox{$\sim$}}}
\def\ltorder{\mathrel{\raise.3ex\hbox{$<$}\mkern-14mu
 \lower0.6ex\hbox{$\sim$}}}

\def\beq{\begin{equation}}
\def\eeq{\end{equation}}
\def\ba{\begin{eqnarray*}}
\def\ea{\end{eqnarray*}}

\newcommand{\et}{{\em et al.}}

\begin{document}

\title{Form factors and radii of light nuclei}

\author{Ingo Sick}
\affiliation{Dept.~f\"{u}r Physik, Universit\"{a}t Basel,
CH4056 Basel, Switzerland}

\date{\today}
\vspace*{5mm}

\begin{abstract}  We discuss the determination of electromagnetic form factors
from the {\em world} data on electron-nucleus scattering for nuclei $Z \leq 3$,
with particular emphasis on the derivation of the moments required for
comparison with measurements from electronic/muonic atoms and isotope  shifts.

\end{abstract}


\email{ingo.sick@unibas.ch}

\maketitle

\section{Introduction}

\noindent The interest in the topic of form factors of light nuclei is threefold: 

1.  The wave functions of light nuclei $A \leq 12$ can be calculated today with
exact methods \cite{Pieper01} starting from the nucleon-nucleon interaction
known from N-N scattering.
Electromagnetic form factors at large momentum transfer $q$ allow for the most
quantitative test of these wave functions including their short-range
properties.

2. Modern experiments provide both charge- and matter isotope shifts for many
stable and unstable nuclei. The radii (rms-radii, Zemach moments) measurable via
electron scattering can provide the reference to convert shifts to absolute
radii. 

3. For the proton there is presently a major discrepancy ($\sim 0.04fm$) between
the rms-radii determined from electron scattering \cite{Sick12} and electronic
Hydrogen \cite{Beyer13} versus   muonic Hydrogen \cite{Pohl10a}. Similar
comparison for light nuclei should provide additional insight.

In this paper we briefly describe the determination of the most precise
electromagnetic form factors. Given the recent
focus on radii, we will place particular emphasis on the determination of the
various moments; for the comparison of the form factors with modern theory we
refer the reader to two review papers \cite{Sick01,Marcucci15}.   

\section{Determination of radii}

Of particular importance for the determination of the rms-radius $R$ of the
nuclear charge density $\rho (r)$ --- which non-relativistically is simply  the
Fourier transform of the monopole charge form factor $G_C(q)$ ---  is the 
question concerning the range of momentum transfers $q$  that is relevant for
the extraction of radii. While the answer "low $q$" is the standard one, the
question is rarely addressed in a quantitative fashion, with the consequence
that improper weight is given to particular sets of data and new experiments
aiming at a determination of R are carried out at uninteresting $q$'s. Using a
standard "notch-test", we have studied this question in detail. In fig.1 we show
the result for the particular case of the proton. Only the data $0.5 < q < 1.2
fm^{-1}$ are really sensitive to the rms-radius. At lower $q$, the finite-size
effect in the form factor  $G(q) \sim 1 - q^2 R^2/6 + ...$ is too small as
compared to the uncertainties of the data, at $q > 1.2 fm^{-1}$ the higher
moments such as $\langle r^4 \rangle$ dominate. Note that for nuclei, which all
have rms-radii larger than the one of the  proton, the maximum sensitivity
occurs at correspondingly lower $q$, {\em i.e.} where $q^2 R^2$ is of similar
size.

The first Zemach moment \cite{Friar04} $\langle r \rangle_{(2)}$, which depends
on both the magnetic and the charge form factor, is needed to predict HFS in
atoms. The third Zemach moment \cite{Friar05b} $\langle r^3 \rangle_{(2)}$ is
needed to relate  the Lambshift in muonic atoms to the  rms-radius. For both
moments, the range of sensitivity is quite similar to the one for the rms-radius
\cite{Sick14a}.

\begin{figure}[bht] \begin{center} 
\includegraphics[scale=0.60,clip]{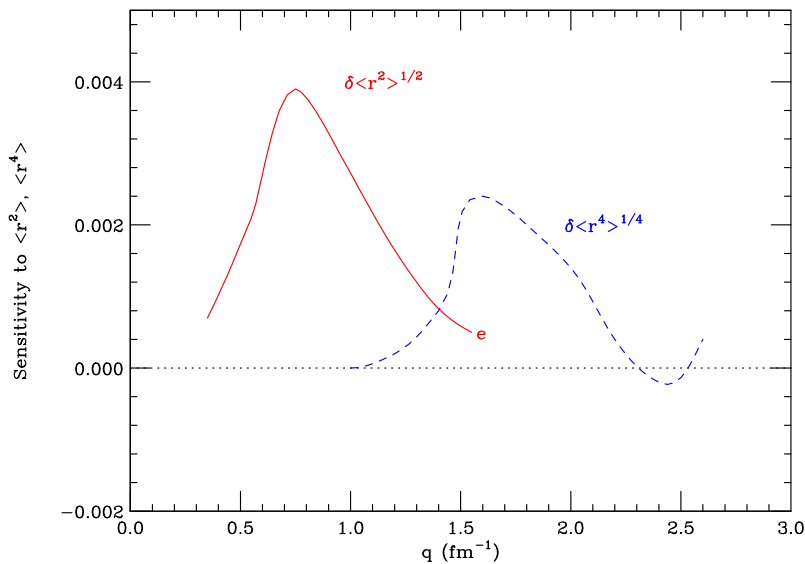} 
{\caption[]{Sensitivity of the
(e,e) data to the lowest moments  $\langle r^2 \rangle$ and $\langle r^4
\rangle$ for the proton. Plotted is the change of the moment for a 1\% change of
the world data in a narrow region $\Delta q$ around a given value of $q$.}}
\end{center}  \end{figure}

A second --- and rather recent --- insight needs to be considered \cite{Sick14}.
The standard procedure of extrapolating $G(q)$ to $q=0$, where the rms-radius
can be obtained from the slope $dG/dq^2$, is unreliable. This extrapolation
requires a parameterization of $G(q)$ which is fitted to the data. Simply taking
a convenient form of $G(q)$ can lead to the situation that the corresponding
$\rho (r)$ has an unphysical behavior at large radii $r$, {\em e.g.} can be
large or not approach zero as we expect from our physics understanding of bound
systems. Such a  behavior at large $r$ can correspond to unreasonable 
moments given the large weight of large $r$ in the moment-determining integrals.
In $q$-space this behavior amounts to an unphysical curvature of $G(q)$
which affects the (implicit) extrapolation from the $q$-region sensitive to R
(see above) to $q=0$ where $R$ is extracted. 

The only way to avoid this problem is to make sure that the large-$r$ density
behaves in a way that is consistent with our  understanding of the physics of
densities. At large $r$, the density of any composite system behaves like a
Whittaker function governed by the separation energy of the lightest charged
constituent \cite{Plattner73}. This Whittaker function (with corrections of minor numerical
impact) can easily be calculated and used to constrain the large-$r$ shape of
$\rho (r)$; typically, this constraint can be used in the region where $\rho
(r)$ has fallen to less than 1\% of its central value. Imposing this constraint
during a fit of the (e,e) data is practical if parameterizations are chosen that
have analytical Fourier transforms so that data (in $q$-space) and constraint
(in $r$-space) can be simultaneously fitted.

The need for constraining the large-$r$ behavior becomes even more apparent once
one considers the integral $\int \nolimits_0^{R_0} \rho (r)~ r^4 dr$ which
determines $R$ in the limit  $R_0 \rightarrow \infty$. In order to
get 98\% of $R$ for {\em e.g.} the deuteron, one has to integrate out to $7fm$!
It is immediately obvious that electron scattering cannot in a significant way
determine the finite size contribution to the form factor at the correspondingly
low $q$ of $\sim 0.2 fm^{-1}$; accordingly, a determination of $R$ to
percent-type accuracy would be illusionary. The usual ''solution'' for this
problem, the use of parameterized form factors, introduces an arbitrariness
(model dependence) that can be avoided only by introducing the physics
constraint as discussed above.  

\section{Deuteron}

The determination of form factors for the I=1 nucleus deuteron is complicated by
the need to separate the three formfactors $G_C$, $G_M$ and $G_Q$. Charge (C+Q)
and magnetic (M) contributions can be separated via forward- and backward-angle
data at the same $q$. Charge monopole (C) and quadrupole (Q) contributions can
be separated if tensor polarization observables, $T_{20}$ in particular, are
available. For the deuteron the data base is quite extensive, with some 450 data
points in the $q$-region $0.2 < q < 10 fm^{-1}$. A fit with the very flexible
Sum-Of-Gaussians (SOG) parameterization \cite{Sick74} of the data (after
correction for Coulomb distortion \cite{Sick98}) supplemented by the large-$r$
tail constraint yields the form factors with error bars $\delta G(q)$ that
account for both the random and the systematic errors of the data. The detailed
comparison of the resulting form factors with modern theory has been discussed
in \cite{Sick01}.

 The charge rms-radius resulting from this fit is given in table 1 and compared 
to the {\em preliminary} value from muonic deuterium, measured by the CREMA
collaboration \cite{Pohl11}, and the radius that can be derived from the
accurate knowledge of the triplet n-p scattering length \cite{Klarsfeld86}.  We
find excellent agreement within the error bar of $0.010 fm$ of the electron
scattering result. This agreement is particularly relevant with regard to the
"proton radius puzzle" as for the proton the discrepancy between (e,e) and
$\mu$H amounts to a significantly larger $\sim 0.04fm$. 

 \begin{table}[htb] 
\begin{tabular}{l|l}
\hline 
(e,e)~~~~ & $~~2.130~ \pm 0.010~ fm$ \\ 
$\mu$H & $~~2.1289 \pm 0.0012 fm$ (prelim.) \\ 
$a_{n-p}$ & ~~$2.131 fm$ \\ 
\hline 
\end{tabular} 
\caption{Deuteron charge rms-radii. The $\mu H$-value is preliminary due to
potential incompleteness of the nuclear polarization correction.} 
\end{table}   

 \section{Helium} 

Electron scattering data are available for $^3$He (~310 data points in the
region $0.2 < q < 10 fm^{-1}$) and $^4$He (190 data points for $0.2 < q < 8.8 fm^{-1}$),
the data for $^4$He being significantly more precise. They are also simpler to
analyze as no error-enhancing C/M-separation is needed. 

The $G_C$ and $G_M$ form factors have been determined as described above for the
deuteron.  A detailed comparison to modern theory is discussed in a review paper
\cite{Sick01}. Overall, it can be stated that the agreement of calculated form
factor and experiment is amazingly good, despite the fact that the form factors
contain substantial contributions due to Meson Exchange Currents MEC. As an
illustration we show in fig.2 the  data for  $^3$He, together with
modern calculations that solve the Schr\"odinger equation for NN potentials
derived from NN scattering. 

\begin{figure}[htb]
 %Figur mit topp/sideways hergestellt, modif boundingBox: 66 206 516 556
%\includegraphics[scale=0.43,clip]{/usr/users/sick/exalph/hfit/dfche3full.ps}
%~~~~~~~~\includegraphics[scale=0.43,clip]{/usr/users/sick/exalph/hfit/dfmhe3full.ps}
\includegraphics[scale=0.43,clip]{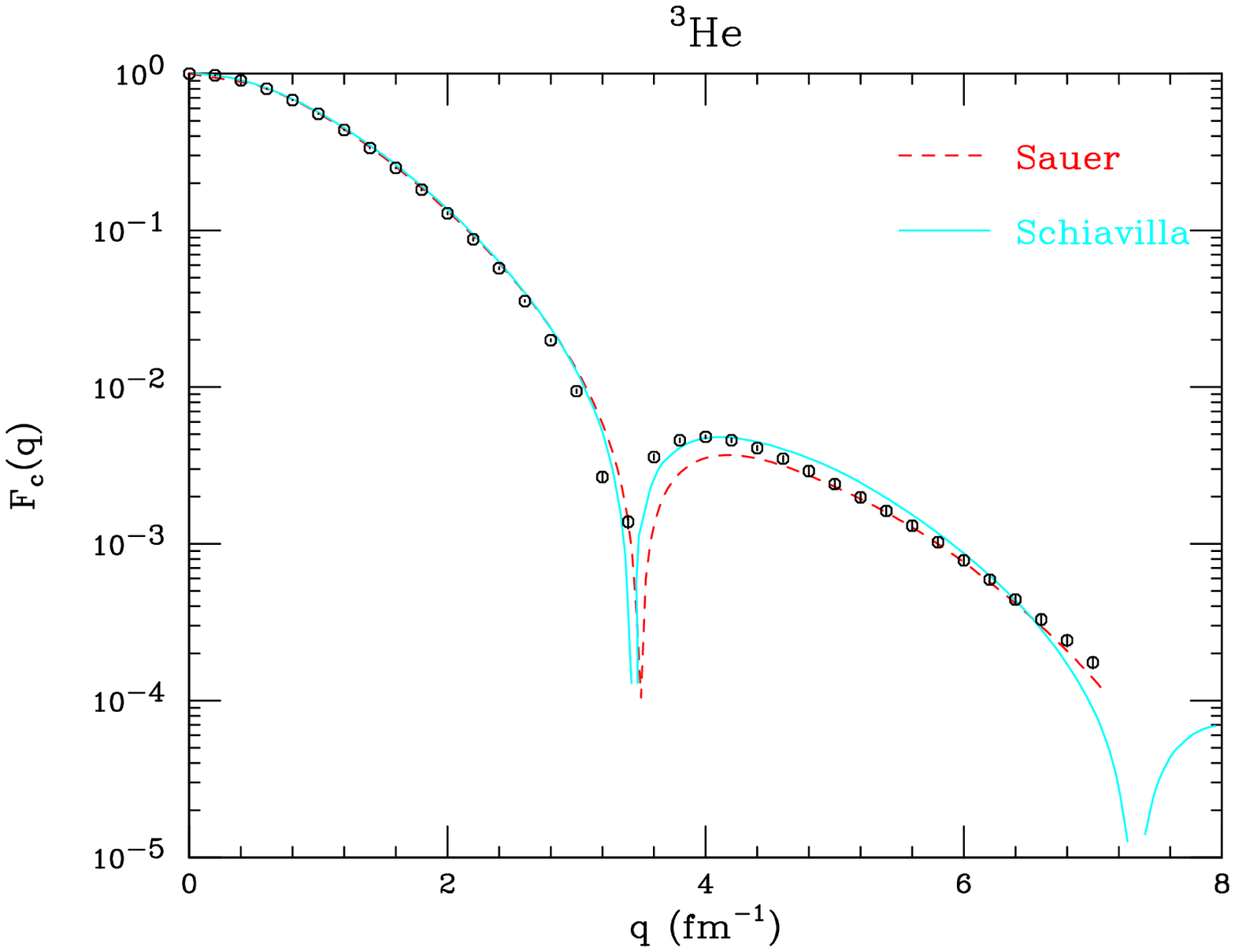}
~~~~~~~~\includegraphics[scale=0.43,clip]{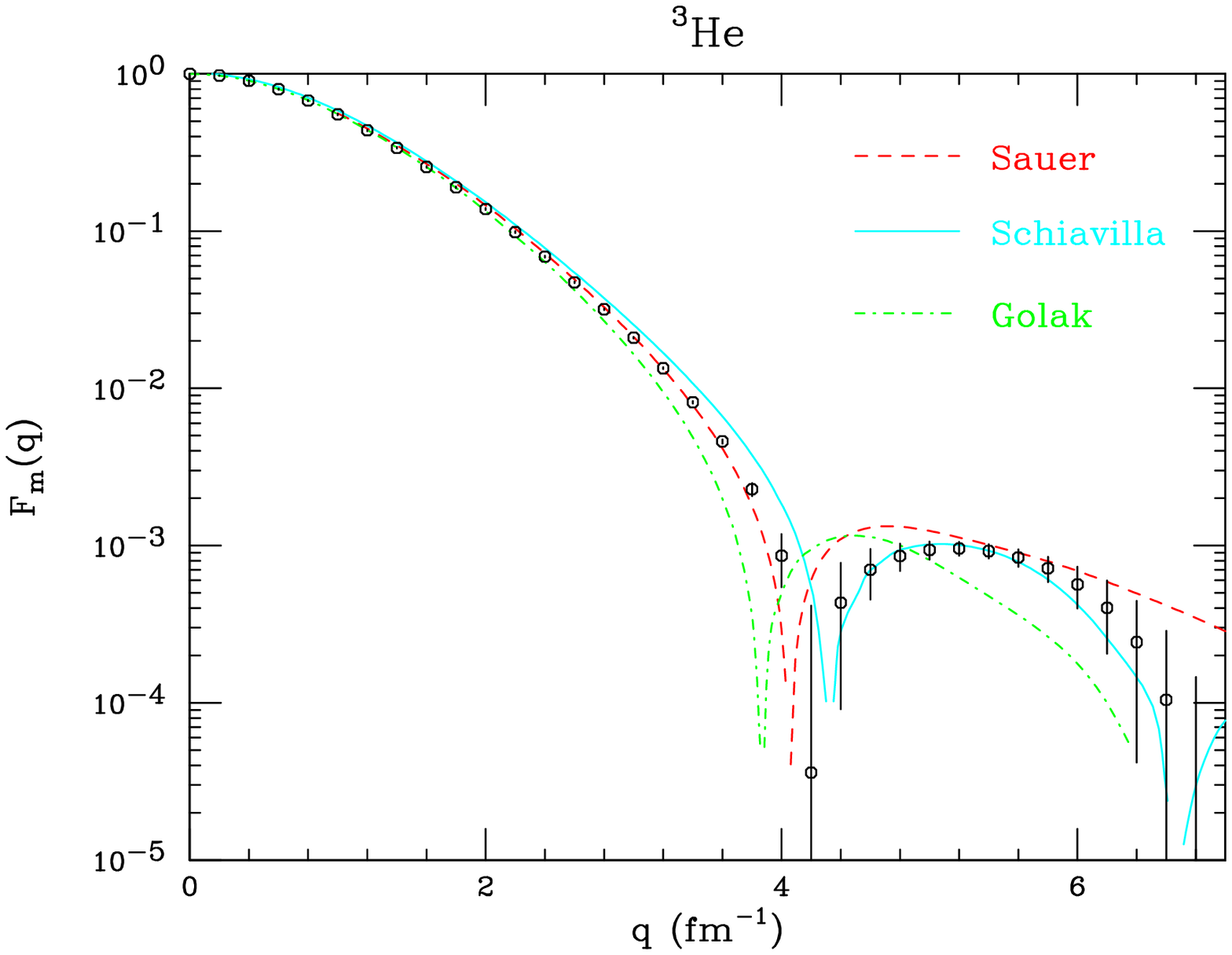}
{\caption[]{Charge and magnetic form factors of $^3$He (points with error bars) 
together with calculations (IA+MEC) of P. Sauer, R. Schiavilla and J. Golak 
\cite{Schiavilla89,Strueve87,Golak01}.}} 
\end{figure}

The rms-radii together with the Zemach moments have been calculated 
\cite{Sick14a} from these
fits and are listed in table 2.
\begin{table}[htb]
\begin{tabular}{l|l|l}
& ~~~~~~~$^3$He  & ~~~~~~~$^4$He  \\
\hline
 rms-radius charge & ~$1.973 \pm 0.016 fm$~ & ~$1.681 \pm 0.004 fm$ \\
 rms-radius magnetism~ & ~$1.976 \pm 0.047 fm$~ & \\ 
$\langle r \rangle_2$ & ~$2.628 \pm 0.016 fm$~  &  \\
 $\langle r^3 \rangle_2$ & ~$28.15 \pm 0.70 fm^3$~ & ~$16.73 \pm 0.10 fm^3$ \\
\hline
\end{tabular}
\caption{Moments of Helium isotopes from (e,e).}
\end{table}

Isotope shifts for the unstable isotopes $^6$He and $^8$He have also been
measured via optical transitions in nuclei stored in traps
\cite{Mueller07,Wang04}, see table 3.   For $^3$He several measurements of the
3-4 isotope shift from optical transitions are available 
\cite{VanRooij11}-\nocite{Pastor12,Pachucki12}\cite{Shiner95}, but they scatter
by roughly the error bar of the electron scattering value.  The  isotope shift
for $^3$He is positive due to the more extended proton wave function resulting
from the lower proton separation energy. Interestingly, the shift for $^8$He is
smaller than the one for $^6$He. Due to the more symmetric configuration of the
extra neutrons --- closed $p_{3/2}$-shell for $^8$He --- the center-of-mass
movement of the $^4$He core in $^8$He has a lesser amplitude and this leads to a
lesser ''smearing'' of the charge. 

Isotope shifts of the matter radii have
been deduced via scattering of GeV/nucleon nuclei on hydrogen in inverse
kinematics \cite{Egelhof02}. They are also listed in table 3.

\begin{table}[htb]
\begin{tabular}{l|l|l}
      & ~~~~$\delta \langle r^2 \rangle_{charge}$ & ~~~$\delta R_{matter}$ \\
\hline
$^3$He~ & ~$1.066 \pm 0.060 fm^2$~   &               \\
$^6$He~ & ~$1.047 \pm 0.034 fm^2$~ &  ~$ 0.81 \pm 0.08 fm$        \\
$^8$He~ & ~$0.911 \pm 0.095 fm^2$~ &   ~$ 0.96 \pm 0.08 fm$       \\
\hline
\end{tabular}
\caption{Charge and matter isotope shifts $^x$He-$^4$He relative to $^4$He. }
\end{table}  

The reference point for all the Helium radii is $^4$He. For this nucleus the
data from electron scattering is the most precise, see fig.3. In
addition, we know for this nucleus not only the {\em shape} of the large-radius
density, but also the absolute value. The extensive set of proton-$^4$He elastic
scattering data has been analyzed using Forward Dispersion Relations FDR
\cite{Plattner73}; the residuum of the singularity closest to the physical
region, due to exchange scattering at $0^\circ$, yields the asymptotic norm of
the proton wave function to $\pm$10\%. This knowledge helps to extract a very
precise charge radius of $1.681 \pm 0.004 fm$ \cite{Sick08} .

\begin{figure}[htb]
%Figur mit topp/sideways hergestellt, modif boundingBox: 66 206 516 556
%\includegraphics[scale=0.5,clip]{/usr/users/sick/exalph/hfit/he4004.rat.ps}
\includegraphics[scale=0.5,clip]{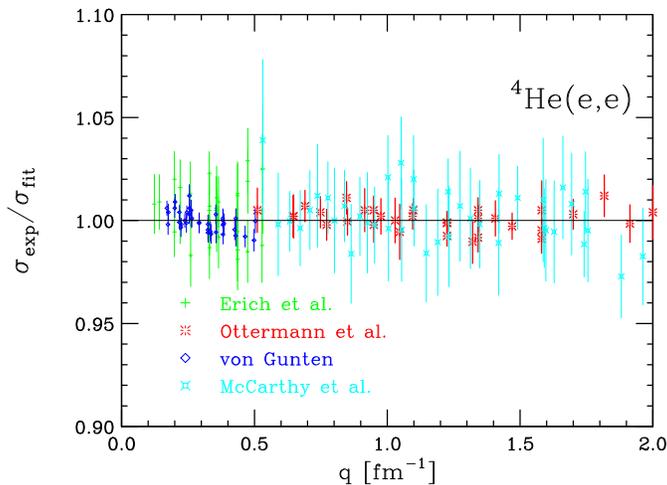}
{\caption[]{Ratio of data/fit for $^4$He at low $q$. The data are from refs.
\cite{Erich68,McCarthy77,vonGunten82,Ottermann85}.}}
\end{figure}

This $^4$He charge rms-radius actually is the most precise radius for any
nucleus determined by electron scattering. It therefore is particularly
interesting to compare this radius to the one being determined from muonic
Helium \cite{Antognini11}. The final value from this experiment is not yet
available, but it is known to be well within the error bar of $\pm 0.004fm$ of
the radius from electron scattering. This agreement of better than $0.004 fm$
is highly significant given the $0.04 fm$ discrepancy between (e,e) and $\mu$X
for the proton.    This speaks strongly against a potential difference between
the ''electromagnetic'' interaction of electrons and muons, an idea  that had
been speculated about in order to explain the proton radius puzzle.

\section{Lithium}

For Lithium electron scattering data are available for A=6,7, the latter being
less accurate and less extensive; moreover they contain both contributions from
the unresolved scattering to the first excited state of $^7$Li and a large
contribution from quadrupole scattering. $^6$Li is therefore the natural
reference nucleus. 

The $^6$Li charge form factor and rms-radius has been determined as discussed
above \cite{Noertershaeuser11}; the data set comprises 86 points in the range  
$0.1 < q < 3.8 fm^{-1}$. A potential complication arises due to the partial
$\alpha + d$ cluster structure and the low deuteron removal energy. For this
case, the asymptotic behavior may be more complex. For the shape of the tail we
have therefore used the one from the GFMC-calculation of Pieper \et
\cite{Pieper01}. As this calculation corresponds to an exact solution of the
Schr\"odinger equation and reproduces the experimental binding energy,   the
large-r shape can be trusted. The resulting charge rms-radius amounts to $2.589
\pm 0.039 fm$. The comparatively large uncertainty results from the limitations
in the (e,e) data.

The charge form factor for $^6$Li agrees very well with the GFMC calculation
\cite{Carlson14} while for the M1 form factor, which also has been measured up to $q \sim 2.8
fm^{-1}$, a small disagreement is observed at the highest $q$'s.

Isotope shifts of the charge rms-radii have been measured by N\"ortersh\"auser
\et ~ (for references see  \cite{Noertershaeuser11}) for mass number up to A=11.
$^{11}$Li is a particularly interesting case of a {\em Borromean} nucleus, as
all  sub-systems, $^{9}$Li+n and n+n (di-neutron), are  unbound while the 3-body
system $^9$Li+n+n is bound. These shifts have been measured at CERN using laser
spectroscopy with the Li-nuclei stored in traps. Fig.4 shows the resulting
radii, together with a series of theoretical calculations discussed in detail in
\cite{Noertershaeuser11}. The unusual behavior of the radii is related to the
fact that $^6$Li and $^7$Li show pronounced effects of cluster structure while
$^8$Li and $^9$Li are closer to a mean-field description.  

 \begin{figure}[htb]
%Figur mit topp/sideways hergestellt, modif boundingBox: 66 206 516 556
%\includegraphics[scale=0.8,clip]{/usr/users/sick/exalph/hfit/lishift.ps}
\includegraphics[scale=0.8,clip]{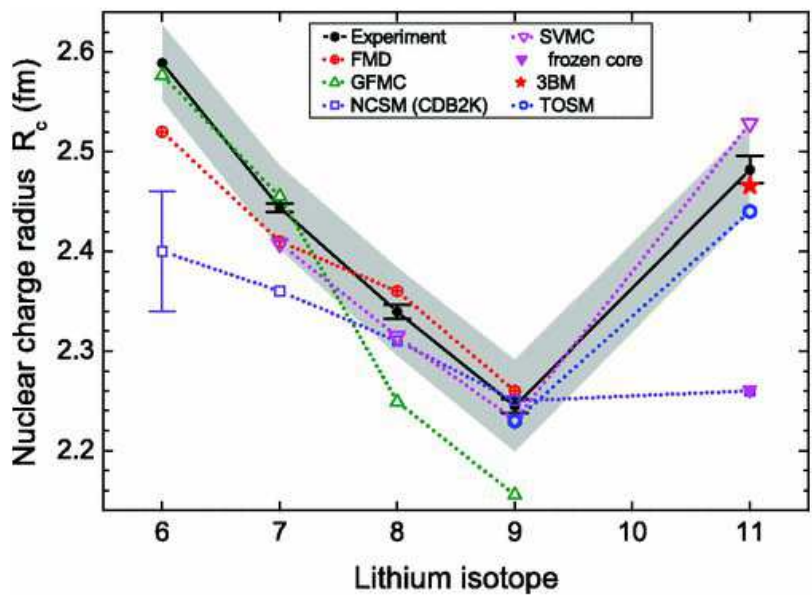}
{\caption[]{Charge radii of Li-isotopes (reference point $^6$Li) from 
\cite{Noertershaeuser11}. The data are from refs.
\cite{Ewald04,Sanchez06,Bushaw03}.}}
 \end{figure}

The matter radii of the Li-isotopes have  been measured at GSI
\cite{Dobrovolsky06}. Experimentally these quantities are accessible (although
with lesser precision)  by scattering of Lithium nuclei, produced by
fragmentation with energies of the order $GeV$/nucleon, from Hydrogen; cross
sections in this energy range can be  interpreted using Glauber theory.  The
resulting data show a more or less constant matter radius throughout the series
A=6--9, with a 30\% increase for A=11 resulting from the extremely low
two-neutron separation energy of only 370$keV$.

\section{Magnetic form factors}

Magnetic scattering for most nuclei is dominated by the properties of valence
neutrons or protons, thus providing an observable governed by different physics. In
general, magnetic form factors also receive larger contributions from meson
exchange currents which, depending on the research goal, is a benefit or a
complication. 

We have in the discussion given above placed little emphasis on these magnetic
form factors. Magnetic form factors are more difficult to determine as
they require experiments at very large scattering angle. As a consequence, the
data base is more limited. A review paper on magnetic
form factors  has been published in \cite{Donnelly84}.  

Magnetic rms-radii are particularly difficult to determine as at low $q$ the
contributions from charge scattering dominate the electron scattering cross
sections. Charge scattering can be suppressed by measurements at 180$^\circ$
scattering angle; since a number of years, however,  no such 180$^\circ$
facilities are available anymore.

\section{Summary}

Precise rms-radii from electron scattering are of interest for comparison with 
radii obtained from modern {\em ab initio} calculations and measurements from
muonic X-ray data; they also serve as an anchor point for the many isotope
shifts now measurable for unstable nuclei. In this paper, we have used the {\em
world} data on electron scattering to determine accurate radii and Zemach
moments for $Z \leq 3$. The uncertainties of the radii derived from (e,e) are
between 0.5-1.5\%. For the special case of $^4$He, where the highest accuracy of
$\sim$0.25\% is reached, we find good agreement with the radius from muonic
Helium, contrary to the case of the proton \cite{Arrington15}  where a 4\% discrepancy  remains.

%\bibliographystyle{unsrt}
%\bibliography{/usr/users/sick/sum3}

\begin{thebibliography}{10}

\bibitem{Pieper01}
S.C. Pieper and R.B. Wiringa.
\newblock {\em Ann. Rev. Nucl. Part. Sci.}, 51:53, 2001.
\vspace*{-2mm}
          
\bibitem{Sick12}
I.~Sick.
\newblock {\em Prog. Part. Nucl. Phys.}, 67:473, 2012.
\vspace*{-2mm}
          
\bibitem{Beyer13}
A.~Beyer {\em et al.}
\newblock {\em Annalen der Physik}, 525:671, 2013.
\vspace*{-2mm}
          
\bibitem{Pohl10a}
R.~Pohl {\em et al.}
\newblock {\em Nature}, 466:213, 2010.
\vspace*{-2mm}
          
\bibitem{Sick01}
I.~Sick.
\newblock {\em Prog. Nucl. Part. Phys.}, 47:245--318, 2001.
\vspace*{-2mm}
          
\bibitem{Marcucci15}
L.E. Marcucci {\em et al.}
\newblock {\em to be publ. in J. Phys. G}.
\vspace*{-2mm}
          
\bibitem{Friar04}
J.L. Friar and I.~Sick.
\newblock {\em Phys. Lett. B}, 579:285, 2004.
\vspace*{-2mm}
          
\bibitem{Friar05b}
J.L. Friar and I.~Sick.
\newblock {\em Phys. Rev. A}, 72:040502, 2005.
\vspace*{-2mm}
          
\bibitem{Sick14a}
I.~Sick.
\newblock {\em Phys. Rev. C.}, 90:064002, 2014.
\vspace*{-2mm}
          
\bibitem{Sick14}
I.~Sick and D.~Trautmann.
\newblock {\em Phys. Rev. C}, 89:012201(R), 2014.
\vspace*{-2mm}
          
\bibitem{Plattner73}
G.R. Plattner {\em et al.}
\newblock {\em Nucl. Phys. A}, 206:513, 1973.
\vspace*{-2mm}
          
\bibitem{Sick74}
I.~Sick.
\newblock {\em Nucl. Phys. A}, 218:509, 1974.
\vspace*{-2mm}
          
\bibitem{Sick98}
I.~Sick and D.~Trautmann.
\newblock {\em Nucl. Phys. A}, 637:559, 1998.
\vspace*{-2mm}
          
\bibitem{Pohl11}
R.~Pohl {\em et al.}
\newblock {\em J. of Physics, Conf. Series.}, 264:012008, 2011.
\vspace*{-2mm}
          
\bibitem{Klarsfeld86}
S.~Klarsfeld {\em et al.}
\newblock {\em Nucl. Phys. A}, 456:373, 1986.
\vspace*{-2mm}
          
\bibitem{Schiavilla89}
R.~Schiavilla, V.R. Pandharipande, and D.O Riska.
\newblock {\em Phys. Rev. C}, 40:2294, 1989.
\vspace*{-2mm}
          
\bibitem{Strueve87}
W.~Strueve {\em et al.}
\newblock {\em Nucl. Phys. A}, 465:651, 1987.
\vspace*{-2mm}
          
\bibitem{Golak01}
J.~Golak {\em et al.}
\newblock {\em Phys. Rev. C}, 63:034006, 2001.
\vspace*{-2mm}
          
\bibitem{Mueller07}
P.~Mueller {\em et al.}
\newblock {\em Phys. Rev. Lett.}, 99:252501, 2007.
\vspace*{-2mm}
          
\bibitem{Wang04}
L.B. Wang {\em et al.}
\newblock {\em Phys. Rev. Lett.}, 93:142501, 2004.
\vspace*{-2mm}
          
\bibitem{VanRooij11}
R.~van Rooij {\em et al.}
\newblock {\em Science}, 333:196, 2011.
\vspace*{-2mm}
          
\bibitem{Pastor12}
P.~{Cancio Pastor} {\em et al.}
\newblock {\em Phys. Rev. Lett.}, 108:143001, 2012.
\vspace*{-2mm}
          
\bibitem{Pachucki12}
K.~Pachucki, V.A. Yerokhin, and P.~Cancio Pastor.
\newblock {\em Phys. Rev. A}, 85:042517, 2012.
\vspace*{-2mm}
          
\bibitem{Shiner95}
D.~Shiner, R.~Dixson, and V.~Vedantham.
\newblock {\em Phys. Rev. Lett.}, 74:3553, 1995.
\vspace*{-2mm}
          
\bibitem{Egelhof02}
P.~Egelhof {\em et al.}
\newblock {\em Eur. Phys. J. A}, 15:27, 2002.
\vspace*{-2mm}
          
\bibitem{Sick08}
I.~Sick.
\newblock {\em Phys. Rev. C}, 77:041302, 2008.
\vspace*{-2mm}
          
\bibitem{Erich68}
U.~Erich {\em et al.}
\newblock {\em Z. Phys.}, 209:208, 1968.
\vspace*{-2mm}
          
\bibitem{McCarthy77}
J.S. McCarthy, I.~Sick, and R.R. Whitney.
\newblock {\em Phys. Rev. C}, 15:1396, 1977.
\vspace*{-2mm}
          
\bibitem{vonGunten82}
A.~von {G}unten.
\newblock {\em Thesis, TH Darmstadt, unpublished}, 1982.
\vspace*{-2mm}
          
\bibitem{Ottermann85}
C.R. Ottermann {\em et al.}
\newblock {\em Nucl. Phys. A}, 436:688, 1985.
\vspace*{-2mm}
          
\bibitem{Antognini11}
A.~Antognini {\em et al.}
\newblock {\em Can. J. Physics}, 89:47, 2011.
\vspace*{-2mm}
          
\bibitem{Noertershaeuser11}
W.~N{\"o}rtersh{\"a}user {\em et al.}
\newblock {\em Phys. Rev. C}, 84:024307, 2011.
\vspace*{-2mm}
          
\bibitem{Carlson14}
J.~Carlson {\em et al.}
\newblock {\em arXiv:1412.3081}, 2014.
\vspace*{-2mm}
          
\bibitem{Ewald04}
G.~Ewald {\em et al.}
\newblock {\em Phys. Rev. Lett.}, 93:113002, 2004.
\vspace*{-2mm}
          
\bibitem{Sanchez06}
R.~Sanchez {\em et al.}
\newblock {\em Phys. Rev. Lett.}, 96:033002, 2006.
\vspace*{-2mm}
          
\bibitem{Bushaw03}
B.A. Bushaw {\em et al.}
\newblock {\em Phys. Rev. Lett.}, 91:043004, 2003.
\vspace*{-2mm}
          
\bibitem{Dobrovolsky06}
A.V.~Dobrovolsky {\em et al.}
\newblock {\em Nucl. Phys. A}, 766:1, 2006.
\vspace*{-2mm}
          
\bibitem{Donnelly84}
T.W. Donnelly and Ingo Sick.
\newblock {\em Review of Modern Physics}, 56:461, 1984.
\vspace*{-2mm}
          
\bibitem{Arrington15}
J.~Arrington and I.~Sick.
\newblock {\em J. Phys. Chem. Ref. Data}, this volume, 2015.
\vspace*{-2mm}
          
      
\end{thebibliography}

\end{document}